\documentclass[format=sigconf]{acmart}

\AtBeginDocument{%
  \providecommand\BibTeX{{%
    \normalfont B\kern-0.5em{\scshape i\kern-0.25em b}\kern-0.8em\TeX}}}

\setcopyright{acmcopyright}
\copyrightyear{2021}
\acmYear{2021}
\acmDOI{10.1145/1122445.1122456}

\acmConference[IFL '21]{33rd Symposium on Implementation and Application of Functional Languages}{September 01--03, 2021}{Online}
\acmBooktitle{IFL '21: ACM Symposium on Implementation and Application of Functional Languages,
  September 01--03, 2021, Online}
\acmPrice{15.00}
\acmISBN{978-1-4503-XXXX-X/18/06}

\usepackage{color}
\usepackage{listings}
\begin{document}

\title{Solving the Funarg Problem with Static Types}

\author{Caleb Helbling}
\email{caleb@purdue.edu}
\orcid{0000-0002-6118-3061}
\affiliation{%
  \institution{Purdue University}
  \streetaddress{610 Purdue Mall}
  \city{West Lafayette}
  \state{Indiana}
  \country{USA}
  \postcode{47907}
}

\author{Fırat Aksoy}
\email{firat.aksoy@ogr.iu.edu.tr}
\orcid{0000-0003-3109-9544}
\affiliation{%
  \institution{Istanbul University}
  \city{Istanbul}
  \country{Turkey}
}

\renewcommand{\shortauthors}{Helbling and Aksoy}

\begin{abstract}
The difficulty associated with storing closures in a stack-based environment is known as the funarg problem. The funarg problem was first identified with the development of Lisp in the 1970s and hasn't received much attention since then. The modern solution taken by most languages is to allocate closures on the heap, or to apply static analysis to determine when closures can be stack allocated. This is not a problem for most computing systems as there is an abundance of memory. However, embedded systems often have limited memory resources where heap allocation may cause memory fragmentation. We present a simple extension to the prenex fragment of System F that allows closures to be stack-allocated. We demonstrate a concrete implementation of this system in the Juniper functional reactive programming language, which is designed to run on extremely resource limited Arduino devices. We also discuss other solutions present in other programming languages that solve the funarg problem but haven't been formally discussed in the literature.
\end{abstract}

\ccsdesc[500]{Software and its engineering~Procedures, functions and subroutines}
\ccsdesc[300]{Software and its engineering~Functional languages}

\keywords{funarg, closure, stack, heap, memory, functional, embedded}

\maketitle

\section{Introduction}
The funarg problem refers to the difficulty associated with compilation of first-class functions in programming languages for stack-based environments. The problem arises in the bodies of nested functions where the nested function refers to identifiers defined in the parent function's lexical scope which are not available in it's own lexical scope. The standard solution is to allocate closures on the heap. There are two variants of this problem: upwards and downwards. The upwards funarg problem refers to the management of the stack when a function returns another function (that may potentially have a non-empty closure), whereas the downwards funarg problem refers to the management of the stack when a function is passed to another function (that may also have a non-empty closure).

Solving the upwards funarg problem is considered to be much more difficult than the downwards funarg problem. The downwards funarg problem is easily solved by copying the closure's lexical scope (captured variables) to the top of the program's stack. For the upwards funarg problem, an issue arises with deallocation. A returning function may capture local variables, making the size of the closure dynamic. Due to this behavior, the closure itself cannot be copied down the stack without causing stack corruption. We must keep in mind that, as soon as the function returns, any local variables will be lost when they are popped off the stack. 

In 1968, Weizenbaum \cite{Weizenbaum1968-zg} made an analysis of function closures using lambda calculus, and showed that function closures in lambda calculus cannot have a stack based data structure and must in fact consist of a tree. Moses \cite{Moses1970-pd} further analyzed the problem for the Algol programming language. Early attempts to solve the funarg problem \cite{Sandewall1971-fw} appear to be hampered by a number of factors, including dynamic scoping, lack of static types and mutable variables.

Despite the titles of these early papers, we contend that the funarg problem (in particular, the upwards funarg problem) has not been adequately solved. In our approach, we solve the upwards funarg problem by statically determining a closure's size at compile time. This allows the closure from the upwards funarg problem to be copied into the calling function, as if it were any other statically sized piece of data. Since the type of a closure's lexical scope can be viewed as a structural record type, we can make our language polymorphic over closures and their lexical scopes. This allows higher order functions such as map (downwards funarg problem) and compose (upwards and downwards funarg problem) to be written very concisely and naturally.

Analyzing closures from a type theoretic perspective is not a new exercise \cite{tc,tcrec,tcsystemf,tccoc}. These papers prefer to treat closures as explicit environments, whose type is determined by an existential. This is not helpful to solving the funarg problem, since existentials are typically treated as boxed values and must be allocated on the heap. Therefore our contribution is unique in the sense that we seek to solve the funarg problem rather than analyzing closures from a purely type theoretic environmental perspective.

\subsection{Contributions}

We describe the following contributions:

\begin{itemize}
    \item We demonstrate how the prenex fragment of System F can be extended to enable stack allocated closures.
    \item We discuss a concrete implementation of this approach in the Juniper programming language.
    \item We analyze the limitations of the extension and identify the connection between types and closures.
    \item We make an analysis of other approaches found in existing programming languages, but haven't been discussed formally in the literature.
\end{itemize}

\section{Extending System F}
In this section, we restrict and extend System F to enable stack allocated closures. We restrict System F to it's prenex fragment which separates regular System F types into type schemes and types. This is a necessary restriction to prevent the emergence of existential types in the type system which prevents determining the sizes of types at compile-time.
The primary addition is that of the closure types, which are defined as function types with their lexical scopes attached to them. The lexical scopes $\delta$ are either type variables or records of captured variables and their respective types. These lexical scope types are are considered equivalent if both the field names and the types of those fields are identical. These additions make it possible for closures to be polymorphic and enable stack allocation of the closure since the size of the closure is known statically at compile-time. Table \ref{tab:grammar} defines the grammar of our language.

\begin{table}[]
\centering
\begin{tabular}{ | c | }
\hline
\\
Expressions\\

$
    e := x \mid \lambda x : A . e \mid e_1\,e_2 \mid \Lambda \alpha . e \mid e\,A
$\\
\\
Type Schemes\\
$
    \sigma := A \mid \forall{a}.\sigma
$\\
\\
Types\\
$
    A, B, C := \alpha \mid A - C \rightarrow B \mid \delta
$\\
\\
Lexical Scope Types\\
$
    \delta := \{ x_0 : A_0, ..., x_n : A_n \}
$\\
\\
Values\\
$   
    v := \left(\lambda x . e\right)\left[ \Delta \right] \mid \Lambda . e
$\\
\\
Contexts\\
$
    \Gamma ::= \emptyset \mid \Gamma,\alpha \mid \Gamma,x:A
$\\
\\
Environments\\
$
    \Delta := \{ x_0 \mapsto v_0, ..., x_n \mapsto v_n \}
$\\\\

\hline
\end{tabular}
\caption{Grammar}
\label{tab:grammar}
\end{table}

\subsection{Big-step semantics}

Stack allocation of closures require the lambda values and their lexical scopes to be propagated and used in pairs. For this reason, we combine the lambda values with their environments into a single value representation. When a lambda is applied, a flat representation of the lambda's lexical scope is constructed and used as the environment along with the lambda's argument for further evaluation. Figure \ref{fig:bigstep} presents the big-step semantics for our language.

\begin{figure}[h]
    \centering
    \begin{equation*}
        \frac
        {v=\Delta(x)}
        {\Delta \vdash x \Downarrow v}
        \hspace{5pt}[\text{\small EVAL-VAR}]
    \end{equation*} \\[5pt]
    
    \begin{equation*}
        \frac
        {}
        {\Delta \vdash \lambda x : A . e \Downarrow (\lambda x . e)[\{y \mapsto \Delta(y) \mid y \in \textbf{fv}(\lambda x : A . e)\}]}
    \end{equation*}
    \begin{equation*}
        [\text{\small EVAL-ABS}]
    \end{equation*}
    \\[5pt]
    \begin{equation*}
        \frac
        {\Delta_1 \vdash e_1 \Downarrow (\lambda x . e_3)[\Delta_2] \quad\; \Delta_1 \vdash e_2 \Downarrow v' \quad\; \Delta_2,x \mapsto v' \vdash e_3 \Downarrow v}
        {\Delta_1 \vdash e_1\,e_2 \Downarrow v}
    \end{equation*}
    \begin{equation*}
        [\text{\small EVAL-APP}]
    \end{equation*}
    \\[5pt]
    \begin{equation*}
        \frac
        {}
        {\Delta \vdash \Lambda \alpha . e \Downarrow \Lambda. e}
        \hspace{5pt}[\text{\small EVAL-TABS}]
    \end{equation*} \\[5pt]
    \begin{equation*}
        \frac
        {\Delta \vdash e_1 \Downarrow \Lambda . e_2\qquad \Delta \vdash e_2 \Downarrow v}
        {\Delta \vdash e_1\,A \Downarrow v}
        \hspace{5pt}[\text{\small EVAL-TAPP}]
    \end{equation*}
    \caption{Big step semantics}
    \label{fig:bigstep}
\end{figure}

\subsection{Typing}

Figure \ref{fig:typingrules} presents the typing rules for our language. There are several changes made to the type rules for the prenex fragment of System F. Of primary interest is the $\small\text{T-ABS}$ rule which also constructs the type of the lexical scope. The construction of the lexical scope itself is performed by $\small\text{T-DELTA}$, which finds all free variables within the lambda, looks up their type in $\Gamma$ and returns the corresponding lexical scope type. The rule $\small\text{T-TABS}$ is also changed to restrict type abstractions to contain no free variables themselves which simplifies the type system and assists in compilation to a stack based environment.

\begin{figure}[h]
    \centering
    \begin{equation*}
        \frac
        {x:A \in \Gamma}
        {\Gamma \vdash x:A}
        \hspace{5pt}[\text{\small T-VAR}]
    \end{equation*} \\[5pt]
    \begin{equation*}
        \frac
        {\Gamma,x:A \vdash e:B \qquad \Gamma \vdash \lambda{x:A}.e \rightsquigarrow \delta}
        {\Gamma \vdash \lambda{x:A}.e : A - \delta \rightarrow B}
        \hspace{5pt}[\text{\small T-ABS}]
    \end{equation*} \\[5pt]
    \begin{equation*}
        \frac
        {\Gamma \vdash e_1: A - \delta \rightarrow B \qquad \Gamma \vdash e_2: A}
        {\Gamma \vdash e_1\,e_2 : B}
        \hspace{5pt}[\text{\small T-APP}]
    \end{equation*} \\[5pt]
    \begin{equation*}
        \frac
        {\Gamma,\alpha \vdash e:\sigma \qquad \textbf{fv}(e)=\emptyset}
        {\Gamma \vdash \Lambda{\alpha}.e:\forall{\alpha}.\sigma}
        \hspace{5pt}[\text{\small T-TABS}]
    \end{equation*} \\[5pt]
    \begin{equation*}
        \frac
        {\Gamma \vdash e: \forall{\alpha}.\sigma}
        {\Gamma \vdash e\,A:\sigma[\alpha:=A]}
        \hspace{5pt}[\text{\small T-TAPP}]
    \end{equation*} \\[5pt]
    \begin{equation*}
        \frac
        {
            y_i \in \mathbf{fv}(\lambda x : A .e) \qquad
            \Gamma \vdash y_0 : C_0, ..., \Gamma \vdash y_n : C_n
        }
        {
            \Gamma \vdash \lambda x : A .e \rightsquigarrow \{ y_0 : C_0, ..., y_n : C_n \}
        }
        \hspace{5pt}[\text{\small T-DELTA}]
    \end{equation*}
    \caption{Typing rules.}
    \label{fig:typingrules}
\end{figure}

\subsection{Examples}

In this section, we will look at two examples that demonstrate that our solution for the funarg problem works for accurately typing two of the most commonly used functions in functional programming: \textbf{map} and \textbf{compose}. For the map example, we will assume that the type system is enriched with a \textbf{list} type constructor. The type of these functions are considerably more noisy than the standard System F types. Fortunately, most of the noise can be elided through type inference.

\begin{equation*}
map : \forall \alpha . \forall \beta . \forall \delta . (\alpha - \delta \rightarrow \beta) - \{\} \rightarrow \alpha\, \mathbf{list} - \{f : \alpha - \delta \rightarrow \beta\} \rightarrow \beta\, \mathbf{list}
\end{equation*}

For the compose example we give the definition of \textbf{compose} as well as its type in our System F extension.

\begin{multline*}
compose := \Lambda \alpha. \Lambda \beta. \Lambda \gamma. \Lambda \delta_1 . \Lambda \delta_2 . \\ \lambda f : (\beta - \delta_1 \rightarrow \gamma) . \lambda g : (\alpha - \delta_2 \rightarrow \beta) . \lambda x : \alpha . f\, (g\, x)
\end{multline*}

\begin{multline*}
compose : \forall \alpha . \forall \beta . \forall \gamma . \forall \delta_1 . \forall \delta_2 . \\ (\beta - \delta_1 \rightarrow \gamma) - \{\} \rightarrow (\alpha - \delta_2 \rightarrow \beta) - \{f : \beta - \delta_1 \rightarrow \gamma\} \rightarrow \alpha \\ - \{f : \beta - \delta_1 \rightarrow \gamma, g : \alpha - \delta_2 \rightarrow \beta\} \rightarrow \gamma
\end{multline*}

\section{Limitations and Extensions}

The most obvious limitation of the extension presented in this paper is that all branches in a program that return a closure must have the exact same lexical scope. One would think that this would be a problem since in most languages, branches of conditionals may return lambdas with different lexical scopes, as long as the types are equivalent. A close analysis of the \textbf{cond} function in System F when combined with Church-encoded Boolean values \textbf{true} and \textbf{false} reveals a surprising result, this problem simply does not appear in the extension described in this paper.

To further elucidate our point, we define \textbf{cond} operator as a function which takes in a true branch and a false branch and a Church-encoded Boolean value (as described in equations  \ref{lambdatrue}, \ref{lambdatruetype}, \ref{lambdafalse}, and \ref{lambdafalsetype}) and returns the respective branch in relation to it's Boolean input.

\begin{multline}
\text{cond} := \Lambda \alpha . \Lambda \beta. \Lambda \gamma. \Lambda \delta \\ \lambda t : \alpha . \lambda f : \beta . \lambda c : (\alpha - \{\} \rightarrow \beta - \delta \rightarrow \gamma) . c\, t\, f
\end{multline}

\begin{multline}
\text{cond} : \forall \alpha . \forall \beta . \forall \gamma . \forall \delta . \\ \alpha - \{\} \rightarrow \beta - \{t : \alpha\} \rightarrow (\alpha - \{\} \rightarrow \beta - \delta \rightarrow \gamma) - \{t : \alpha, f : \beta\} \rightarrow \gamma
\end{multline}

\begin{equation}
\label{lambdatrue}
\text{true} := \Lambda \alpha . \Lambda \beta . \lambda t : \alpha. \lambda f : \beta . t
\end{equation}

\begin{equation}
\label{lambdatruetype}
\text{true} : \forall \alpha . \forall \beta . \alpha - \{\} \rightarrow \beta - \{t : \alpha\} \rightarrow \alpha
\end{equation}

\begin{equation}
\label{lambdafalse}
\text{false} := \Lambda \alpha . \Lambda \beta . \lambda t : \alpha. \lambda f : \beta . f
\end{equation}

\begin{equation}
\label{lambdafalsetype}
\text{false} : \forall \alpha . \forall \beta . \alpha - \{\} \rightarrow \beta - \{\} \rightarrow \beta
\end{equation}

Applying the \textbf{cond} operator with \textbf{true} and \textbf{false} in equations \ref{trueexample} and \ref{falseexample} reveals why Church-encoding does not suffer from the branching problem. In System F, the branch chosen by \textbf{cond} is already present in the type system by selecting the type for $\gamma$. Fundamentally, this occurs because \textbf{true} and \textbf{false} have different type signatures, that also determine the branch taken.

\begin{equation}
\label{trueexample}
((\text{cond}\, \alpha\, \beta\, \alpha\, \{t : a\})\, t\, f\, (\text{true}\, \alpha\, \beta\,)) : \alpha
\end{equation}

\begin{equation}
\label{falseexample}
((\text{cond}\, \alpha\, \beta\, \beta\, \{\})\, t\, f\, (\text{false}\, \alpha\, \beta\,)) : \beta
\end{equation}

However, in a language with sum types and pattern matching (as Boolean types are in most languages), the branching problem with closures will be present. Whether or not this is a major issue is debatable, however, we can say for certain that this removes some generality from our approach. An interesting approach to recovering this generality is a paper on open closures by Scherer and Hoffmann \cite{openclosure}, however their approach only applies to simply typed lambda calculus and not for System F.

The second limitation appears under the presence of mutually recursive functions. When compiling top level functions, we can consider their closure to be empty since we can refer to them using a simple function pointer. For inner functions that are often mutually recursive, the situation is more complicated. The lexical scope of all functions declared in a \textbf{let-rec} block can be constructed by taking the union of all of the lexical scopes of all the functions declared in the block. This constructs a common lexical scope that can be passed to all the related functions. Therefore, all functions declared in the \textbf{let-rec} block will have identical lexical scope types.

When one function in the let-rec block calls another function in the same let-rec block, it simply passes the lexical scope that it was passed. Within the lexical scope type itself, we place all the functions within the let-rec block with the non-closure function type $A \rightarrow B$. This indicates that this function should be passed the lexical scope that this function is contained within.

Extending the system presented in this paper to the impredicative System F at first glance seems feasible. We believe that the funarg problem is orthogonal issue to that of first-class polymorphism and stack allocated existential types. However we have yet to make a full analysis of how the lexical environments presented here interact with System F. One possible issue that may crop up is subtle interactions with existential types, which can be encoded in System F \cite{proofsandtypes}.

\section{Implementation}

We have implemented the closure system given here in Juniper \cite{juniper}, a functional reactive programming language for the Arduino. The Juniper language transpiles all of its code to C++. The analysis of C++ in section \ref{existing} reveals the issues facing the use of lambdas as a transpile target. To work around the issues presented in that section, we implement our own \textbf{function} class. This class is constructed by passing in a lexical scope, along with a function pointer. In C++, lambda types that do not capture anything can be casted to type safe function pointers. We provide our own stack allocated closure support by passing in the lexical scope as the first argument to the underlying lambda function. Within this lambda, the compiler inserts variable assignments which extracts the contents of the lexical scope into local variables. Lambdas that do not capture anything are not passed a lexical scope.

\begin{lstlisting}
template <typename ClosureType, typename Result, typename ...Args>
class function;

template <typename Result, typename ...Args>
class function<void, Result(Args...)> {
private:
    Result(*F)(Args...);

public:
    function(Result(*f)(Args...)) : F(f) {}

    Result operator()(Args... args) {
        return F(args...);
    }
};

template <typename ClosureType, typename Result, typename ...Args>
class function<ClosureType, Result(Args...)> {
private:
    ClosureType Closure;
    Result(*F)(ClosureType&, Args...);

public:
    function(ClosureType closure, Result(*f)(ClosureType&, Args...)) : Closure(closure), F(f) {}

    Result operator()(Args... args) {
        return F(Closure, args...);
    }
};
\end{lstlisting}

A lexical scope struct is defined for every combination of environment names discovered in the program. The types of the variables are made generic, so that these lexical scope structs can be declared before any of the other types in the program. Below is an example of the lexical scope struct generated by the Juniper compiler for the \textbf{compose} function. These structs are created and passed to the \textbf{function} constructor the moment the lambda declared. This ensures that the closure contains a snapshot of the current environment. Importantly, if a mutation of any local variables occurs after a closure has been constructed, the closure will not be updated (since it essentially takes a snapshot of the local environment the moment it is created). For purely functional languages this is not an issue since variables cannot be mutated.

\begin{lstlisting}
// Compose lexical scope
template<typename T1,typename T2>
struct closuret_0 {
    T1 f;
    T2 g;

    closuret_0(T1 init_f, T2 init_g) :
        f(init_f), g(init_g) {}
};
\end{lstlisting}

\section{Existing Approaches} \label{existing}

In this section we will analyze a number of different existing languages and how they solve the funarg problem. To our knowledge the system presented in this paper is the only language that will allow the type of a closure to be fully written out with text (ie, it does not generate an implicit type or class).

In C++, each declaration of a lambda implicitly declares a new class \cite{cppstandard} and therefore type in the system. Since these classes are auto-generated, it is impossible to write their types using text, and programmers must rely on the auto keyword. These C++ lambdas are indeed stack allocated since their underlying representation is a class with a operator() defined. The problems begin however as soon as you start passing around or returning lambdas. Higher order functions that consume lambdas must be made generic over all of these implicit lambda classes. This makes it is impossible to add any sort of constraints on the types of the arguments of the lambda, or the return type of the lambda. Returning a lambda also becomes very annoying, and the auto keyword must be used as the return type. Once again, we see that it is impossible to constrain the return type to anything in particular. The solution used by most C++ programmers is to make use of the standard library std::function, which turns the stack allocated closure into a heap allocated entity. Perhaps in the future, clever use of C++ concepts could add extra type safety to the use of these implicit lambda classes.

To our knowledge, the only other language that has similar capabilities to what is presented in this paper is Rust \cite{rustlang}. Just as in C++, the Rust system operates by generating a unique type for every closure. However Rust goes a step further with its trait system. Like in C++, accepting a closure as a parameter to another function requires the use of a single generic parameter. Unlike C++, the type of this parameter can be constrained with traits. Additionally, returning a closure requires the textual use of one of these traits.

\section{Future Work}

In the future we would like to further extend our type system for enriched languages. We are interested in how adding conditionals and sum types to the language will force changes in the handling of the lexical scopes. We also would like to give a more thorough description of how to handle the let-rec case.

We would also like to extend this system for the impredicative System F without restrictions as much as possible, which we predict to potentially force additional changes in the representation of the lexical scopes.

\section{Conclusion}

In this paper we have described an extension to the prenex fragment of System F that solves the funarg problem. Our solution solves both the downwards and the upwards funarg problem and allows closures to be completely stack allocated. Functional languages can take advantage of this new system to optimize memory usage, and this extension is particularly useful on embedded systems. We also discuss extensions to our language and identify the possibility to extend this work to solve the conditional problem as well as stack allocation for first-class polymorphism. We discussed how existing languages solve the funarg problem, and conclude that our approach is the first so far that allows writing closure type signatures without auto-generated types or classes.




\end{document}